\documentclass[aps, pre, amsmath,amssymb,twocolumn]{revtex4}
\usepackage{graphicx}
\usepackage{bm}
\usepackage{epsfig}
\begin{document}

\title{\bf A non-equilibrium Monte Carlo approach to potential refinement in inverse problems}

\author{Nigel B. Wilding}
\affiliation{Department of Physics, University of Bath, Bath BA2 7AY, United Kingdom}


\begin{abstract} 

The inverse problem for a disordered system involves determining the
interparticle interaction parameters consistent with a given set of
experimental data. Recently, Rutledge has shown (Phys. Rev. {\bf E63},
021111 (2001)) that such problems can be generally expressed in terms
of a grand canonical ensemble of polydisperse particles. Within this
framework, one identifies a polydisperse attribute (`pseudo-species')
$\sigma$ corresponding to some appropriate generalized coordinate of
the system to hand. Associated with this attribute is a composition
distribution $\overline\rho(\sigma)$ measuring the number of particles
of each species. Its form is controlled by a conjugate chemical
potential distribution $\mu(\sigma)$ which plays the role of the
requisite interparticle interaction potential. Simulation approaches to
the inverse problem involve determining the form of $\mu(\sigma)$ for
which $\overline\rho(\sigma)$ matches the available experimental data.
The difficulty in doing so is that $\mu(\sigma)$ is (in general) an
unknown {\em functional} of $\overline\rho(\sigma)$ and must therefore
be found by iteration. At high particle densities and for high degrees
of polydispersity, strong cross coupling between $\mu(\sigma)$ and
$\overline\rho(\sigma)$ renders this process computationally 
problematic and laborious. Here we describe an efficient and robust
{\em non-equilibrium} simulation scheme for finding the equilibrium
form of $\mu[\overline\rho(\sigma)]$. The utility of the method is
demonstrated by calculating the chemical potential distribution
conjugate to a specific log-normal distribution of particle sizes in a
polydisperse fluid.


\end{abstract} 
\maketitle
\setcounter{totalnumber}{10}

\section{Introduction and background}

\label{sec:intro}

Much of statistical mechanics is concerned with the task of deducing the
macroscopic structure of a system, {\em given} a description of its
microscopic interaction interactions. Sometimes, however, one is confronted
with the reverse problem: that of determining the form of the interactions
from knowledge of the structure. This so-called ``inverse problem''
\cite{LEVESQUE85} arises, for example, when one has scattering measurements of
the structure factor of a molecular fluid and wishes to find the corresponding
interparticle interaction potentials \cite{SCATTERING}. Alternatively one
might have obtained spectroscopic data for the distribution of orientations of
a certain bond in a molecular solid and wish to determine the corresponding
bond orientation potential \cite{SPECTROSCOPY}.

One approach of longstanding for tackling inverse problems is Reverse
Monte Carlo \cite{MCGREEVY}. This simulation scheme seeks to minimize
error estimators quantifying the difference between experimentally
derived structural data (such as a radial distribution function) and
the corresponding simulation averages. The minimization proceeds by
replacing the Hamiltonian in the standard MC Metropolis update scheme
with the error estimator function, while the role of temperature is
replaced by the estimated degree of uncertainty in the experimental
data. The method outputs a set of configurations, which are consistent
with the experimental data and can be further analyzed, but provides no
direct information on the underlying form of the interaction potential,
the values of the thermodynamic fields, or fluctuation effects.

Recently, Rutledge \cite{RUT01} has suggested that ostensibly disparate
inverse problems can be incorporated within a unifying theoretical
framework by mapping them onto a generalized grand canonical
polydisperse composition space. The key to achieving this is the
identification of a generalized coordinate of the system, $\sigma$, on
which the potential function of interest is defined. The problem may
then be translated into the language of polydispersity by regarding
$\sigma$ as a continuous variable labeling the species of particles
that comprise the system. In general, however, these polydisperse
particles need not correspond directly to the real particles of the
system, and are perhaps best regarded as `pseudo particles' the precise
identity of which depends on the particular choice of $\sigma$ and thus
on the physical situation to hand. The reformulation is completed by
assuming that the number density of pseudo particles of each $\sigma$
is free to fluctuate. Within the resulting open ensemble, the
instantaneous count of particles of each species is measured by a
composition distribution $\rho(\sigma)$, the ensemble-averaged form
of which, $\overline{\rho}(\sigma)$, is directly controlled by the
conjugate chemical potential distribution $\mu(\sigma)$ (see
Appendix~\ref{append}). The latter quantity plays the role of the
potential function of interest.

In practice a surprising variety of inverse problems can be cast in this form.
For example, in a liquid crystal, one can identify $\sigma$ with the
orientation of a molecule (cf. \cite{ONSAGER}) and $\mu(\sigma)$ with an
effective orientational potential. In an atomic crystal, $\sigma$ can be
identified with the phonon frequencies \cite{RUT01} and $\mu(\sigma)$ to the
Fourier transform of the interparticle potential. For a simple monatomic
fluid, one can construct a ``bond'' picture, in which each of the $N$ real
particles is replaced by $N$ non-interacting pseudo particles, all localized
to the site of the real particle. Each pseudo particles is speciated according
to its distance to one other nominated real particle and the associated
$\mu(\sigma)$ is simply the interparticle potential \cite{NOTE0}.

In general, for systems described by two-body particle interactions, a
unique relationship is believed to exist between
$\overline\rho(\sigma)$ and $\mu(\sigma)$
\cite{HENDERSON,ZWICKER,EVANS}. The simulation challenge is then to
determine the form of $\mu(\sigma)$ for which $\overline\rho(\sigma)$
matches some ``target'' form $\overline\rho_t(\sigma)$ corresponding to
the available experimental structural data. The difficulty in achieving
this is that $\mu(\sigma)$ is an unknown {\em functional} of
$\overline\rho(\sigma)$, and thus must be determined iteratively via
some refinement procedure. In the following section we summarize the
state-of-the-art in this regard.

\section{Iterative refinement schemes and convergence issues}

Refinement typically commences using some guessed form for the
interaction potential $\mu(\sigma)$. This serves as input to a Monte
Carlo simulation, in the course of which the form of
$\overline{\rho}(\sigma)$ is obtained as a histogram. Once sufficient
statistics for $\overline{\rho}(\sigma)$ have been accumulated, the
next and successive iterations modify $\mu(\sigma)$ to reduce the
discrepancy between $\overline{\rho}(\sigma)$ and the target function
$\rho_t(\sigma)$.  To achieve this Rutledge \cite{RUT01} utilized an
iteration scheme based on a zeroth order (i.e. non-interacting or ideal
gas) approximation to the chemical potential distribution: 

\begin{equation}
\mu^{ig}(\sigma)=\ln\left(\overline{\rho}(\sigma)\right)\;,
\end{equation}
which can be viewed as the polydisperse generalization of the
potential of mean force \cite{NOTECP,HANSEN}. Using this approximation
to initialize $\mu(\sigma)$, the iteration then proceeds according to

\begin{equation}
\mu_{i+1}(\sigma)=\mu_i(\sigma) -
\alpha\ln\left(\frac{\overline{\rho}_i(\sigma)}{\rho_t(\sigma)}\right)\;,
\label{eq:pmfa}
\end{equation}
where $\alpha\le 1$ is a step-size parameter, the value of which must
be chosen sufficiently small to ensure that the method converges, but
not so small that convergence is excessively slow. Rutledge applied
this approach to calculate the effective interparticle interaction
potential of a single component fluid consistent with a prescribed
radial distribution function. For this particular purpose, his scheme
is similar to the potential refinement method of Soper \cite{SOPER96}.

The efficiency with which inverse problems can be solved is clearly
contingent on the convergence rate of the refinement procedure for
$\mu(\sigma)$.  Typically a strategy such as that of Eq.~\ref{eq:pmfa}
will be effective at low number densities of pseudo particles where the
non-interacting approximation retains some degree of accuracy.
Difficulties arise, however, at high number densities and for high
degrees of polydispersity (ie. slowly varying forms of
$\overline\rho(\sigma)$). In this regime, $\mu(\sigma)$ can differ
greatly from $\mu^{ig}(\sigma)$. Moreover, strong cross coupling
between $\mu(\sigma)$ and $\overline\rho(\sigma)$ implies that
variations (within the refinement procedure) to the chemical potential
at one value of $\sigma$ can significantly affect the entire form of
$\overline\rho(\sigma)$. This can seriously impede convergence; the
sole remedy being to utilize a small step size $\alpha$, with a
concomitant increase in the computational effort \cite{LYUBARTSEV95}.

In view of these considerations, one might seek to adopt a more
sophisticated refinement procedure in the expectation that it will
promote faster convergence than Eq.~\ref{eq:pmfa}. Such schemes have
been developed and applied by Lyubartsev \& Laaksonen
\cite{LYUBARTSEV95} and T\'{o}th \cite{TOTH}.  They employ large
matrices of derivatives (obtainable via fluctuation relations as
ensemble averages) and designed to direct the minimization more
reliably in the `downhill' direction in the multidimensional parameter
space of some cost function measuring the deviation of
$\overline\rho(\sigma)$ from $\rho_t(\sigma)$. Whilst in our experience
of gradient schemes (we have tried Powell's method, conjugate gradient
methods and variable metric methods \cite{NUMREC}), they do provide a
more rapid approach towards the solution in the early iterations, we
find that they do not converge as reliably (at least for high degrees
of polydispersity) as the non-interacting approximation scheme of
eq.~\ref{eq:pmfa}. Instead we have found them to be prone to becoming
trapped in local minima of the cost function. Although, empirically,
this problem can be ameliorated by mixing iterations \cite{NOTE1} of
the sophisticated scheme with simple ones of the form
Eq.~\ref{eq:pmfa}, such an {\em ad-hoc} approach is clearly neither
theoretically well-founded nor aesthetically satisfying.

Other recent work, carried out in the context of targeted
polydispersity, has examined the utility of histogram extrapolation in
targeting a specific composition distribution $\rho_t(\sigma)$
\cite{ESCOBEDO,WILDING02}. Histogram extrapolation \cite{HR} allows a
histogram for $\overline\rho(\sigma)$ accumulated at some $\mu(\sigma)$
to be reweighted to provide an estimate for $\overline\rho(\sigma)$
corresponding to some other chemical potential distribution
$\mu^\prime(\sigma)$. To exploit the method, it can be embedded within
a gradient-based scheme to minimize a cost function measuring the
deviation of $\overline\rho(\sigma)$ and $\rho_t(\sigma)$
\cite{WILDING02}. In practice, however, one finds that the
extrapolation operates effectively only within a limited range of
chemical potential space around the simulation state point. If one
attempts to extrapolate too far from this point, spurious structure
appears in the results and the method may not converge to the correct
solution. This problem can be dealt with by constructing a series of
intermediate targets that interpolate between the initial guess for
$\overline\rho(\sigma)$ and the true target (see ref.~\cite{WILDING02}
for full details).

In addition to the convergence issues already mentioned, a further
important matter is the statistical quality of the measured estimate
for $\overline\rho_i(\sigma)$. Unless the `signal-to-noise' ratio
associated with this measurement is sufficiently high prior to each
iteration, the procedure will necessarily fail to converge. Often,
however, one has no ready criteria for knowing when the statistical
quality of the data is ``good enough''. We note also in passing that
related problems can arise when the statistical quality of the
experimentally derived target distribution $\rho_t(\sigma)$ is poor.
Indeed, it has been reported that in such circumstances the solution to
which the refinement converges can depend on the initial guess for
$\mu(\sigma)$ \cite{SOPER1} or contain spurious features \cite{TOTH1}.
This would seem to suggest that `noise' in the target can engender
additional local minima in the multidimensional parameter space on
which the cost function is defined.

In this paper we describe a new iterative refinement scheme which is
simple and straightforward to implement, yet yields a robust and
efficient procedure. Our method (which is inspired by, but is distinct
from, the flat energy histogram random walk algorithm of Wang and
Landau \cite{WL}) is based on a non-equilibrium MC procedure in which
refinements to $\mu(\sigma)$ are implemented `on-the-fly' during the
simulation, not between iterations as in equilibrium schemes.
Empirically we find that the method provides an estimate of the true
solution right from the earliest iterations; subsequent iterations
merely serve to reduce the statistical uncertainty in this estimate.
This attractive feature promotes more reliable convergence than is
found in more complicated gradient based schemes. In a comparative
test, the new method is found to be significantly more efficient than
the simple equilibrium iteration scheme described by eq.~\ref{eq:pmfa}.

\section{Non-equilibrium refinement scheme}

We consider systems having a polydisperse attribute $\sigma$ varying
continuously in the range $0\le\sigma \le\sigma_c$. The associated
composition distribution $\rho(\sigma)$ (defined in appendix~\ref{append})
and its conjugate chemical potential distribution are represented as
histograms formed by binning the $\sigma$-domain into an integer number
$M$ of sub-intervals. A discussion of this binning strategy can be
found in ref. \cite{WILDING02}. 

Given some prescribed target histogram $\rho_t(\sigma)$,  the
refinement  $\mu(\sigma)$ proceeds iteratively using MC simulation. In
general, no special bootstrapping measures are necessary for
initializing $\mu(\sigma)$ and, in the absence of a better guess, one
can simply take $\mu(\sigma)=C\hspace*{1mm} \forall\hspace*{1mm}
\sigma$, with $C$ an arbitrary constant.  During the course of an
iteration, the chemical potential distribution is modified at 
regular short intervals (eg. every 10 MC sweeps). The modification
continuously tunes $\mu(\sigma)$ such as to minimize the deviation of
the {\em instantaneous} composition distribution $\rho(\sigma)$ from
its target value:

\begin{equation}
\mu^\prime(\sigma)=\mu(\sigma)- \beta_i\left(\frac{\rho(\sigma)-\rho_t(\sigma)}{\rho_t(\sigma)}\right)\hspace*{2mm}\forall\hspace*{1mm}\sigma\:,
\label{eq:modify}
\end{equation}
where $\beta_i$ is the modification factor for the $i$th iteration. 
Updating $\mu(\sigma)$ `on-the-fly' in this manner quickly reduces
differences between $\rho(\sigma)$ and its target \cite{NOTE3}. 

The criterion by which an iteration is deemed to have completed is
based on the maximum value, across the range of $\sigma$, of the
relative deviation of the {\em average} composition distribution from the target form:

\begin{equation}
\zeta={\rm max}\:\left( \left| \frac{\overline\rho_i(\sigma)-\rho_t(\sigma)}{\rho_t(\sigma)}\right|\right)\:.
\label{eq:zeta}
\end{equation}
Once $\zeta$ falls below some previously specified threshold value $\zeta^*$,
$\mu(\sigma)$ can be saved and used as the starting
point for the second iteration.  This proceeds similarly to the first iteration, except
that the value of the modification factor $\beta$ is reduced, viz

\begin{equation}
\beta_{i+1}=\beta_i/n\:,
\end{equation}
with $n$ some (small) integer \cite{NOTE4}.

Successive further iterations steadily reduce $\beta$ towards zero, thereby
restoring detailed balance and driving $\mu(\sigma)$ towards its equilibrium
limiting form. For sufficiently large $i$ (i.e. sufficiently small $\beta_i$),
$\mu_i(\sigma)$ provides a good approximation to this limiting form. If desired,
however, one can extrapolate fully to the equilibrium limit by performing a
final run with $\beta=0$, followed by application of histogram extrapolation
\cite{HR,WILDING02} to $\mu(\sigma)$ in order to match $\overline\rho(\sigma)$
to $\rho_t(\sigma)$.

\section{Results}

In order to gauge the efficacy of our scheme we have tested it on a
challenging problem: namely that of determining the chemical potential
distribution for a fluid of polydisperse Lennard-Jones (LJ) particles
having an extremely broad distribution of sizes. 

The interparticle potential between two particles labeled $i$ and $j$
is given by

\begin{equation}
u(r_{ij})=4\epsilon_{ij}\left[\left(\frac{\sigma_{ij}}{r_{ij}}\right)^{12} - \left(\frac{\sigma_{ij}}{r_{ij}}\right)^6\right]\:,
\label{eq:plj}
\end{equation}
with $\sigma_{ij}$ given by the additive mixing rule
\begin{equation}
\sigma_{ij}=(\sigma_i+\sigma_j)/2.
\label{eq:mix}
\end{equation}
The target distribution of particle diameters $\sigma$ appearing in eq.~\ref{eq:mix} was
assigned a log-normal form, described by the normalized shape function:
\begin{eqnarray}
f(\sigma)&=&\frac{1+W^2}{\overline{\sigma}\sqrt{2\pi\ln(1+W^2)}}\nonumber\\
&\times& \exp\left(-\frac{[\ln(\sigma/\overline{\sigma})+(3/2)\ln(1+W^2)]^2}{2\ln(1+W^2)}\right).
\label{eq:lognorm}
\end{eqnarray}
Here $\overline{\sigma}$ is the average particle diameter and $W$ is
the standard deviation in units of $\overline{\sigma}$. The associated
target composition distribution follows as $\rho_t(\sigma) = \rho_t\,
f(\sigma)$, where $\rho_t$ is the target number density. 

For computational convenience, $f(\sigma)$ was truncated at $\sigma_c=12$. This
choice implies that the largest permitted particle size has a volume $1728$
times that of the average particle size ($\overline{\sigma}=1$). The
simulations were performed within a periodic cubic box of side
$L=90\overline\sigma$ and the temperature was set to $T=2.35$ (in standard LJ
units \cite{FRENKEL}). The width and scale of the target polydispersity
distribution were set to $W=2.5$ and $\rho_t=0.0152$ respectively,
corresponding to a target volume fraction, $\eta\approx 13\%$ where

\begin{equation}
\eta\equiv\int_0^{\sigma_c} d\sigma \frac{\pi}{6}\sigma^3\overline\rho(\sigma)\;.
\end{equation}
Although this volume fraction would not be considered especially high
for a pure fluid (the critical volume fraction for a monodisperse LJ
fluid is about $16\%$), it is sufficient in the present context to
ensure a high degree of cross coupling between $\overline\rho(\sigma)$
and $\mu(\sigma)$. 

The grand canonical ensemble Monte Carlo (GCMC) algorithm employed has
a Metropolis form \cite{FRENKEL} and invokes four types of operation:
particle displacements, particle insertions, particle deletions and
particle resizing; each is attempted with equal frequency. Specific to
the polydisperse case is the resizing operation which entails
attempting to change the diameter of a nominated particle by an amount
drawn from a uniform random deviate constrained to lie in some
prescribed range. This range (maximum diameter step-size) is chosen to
provide a suitable balance between efficient sampling and a
satisfactory acceptance rate at the prevailing number density. As
regards the remaining types of moves, these proceed in a manner similar
to the monodisperse case \cite{FRENKEL}, except that for insertion
attempts the new particle diameter is drawn with uniform probability
from the range $\sigma \in [0,\sigma_c]$. 

\begin{figure}[h]
\includegraphics[width=8.5cm,clip=true]{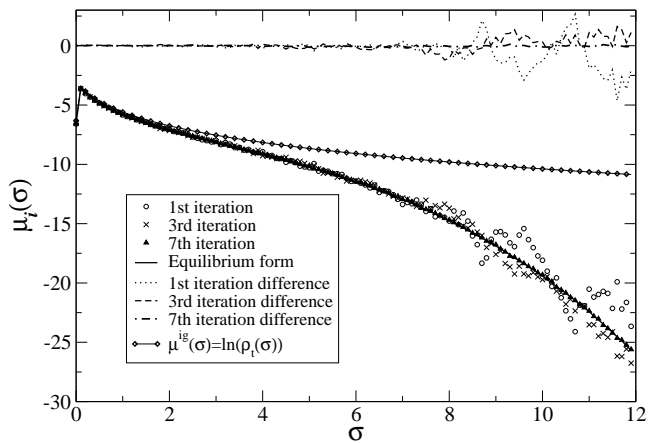}
\caption{The convergence of the non-equilibrium refine procedure. Shown
(bottom part) are the forms of $\mu_i(\sigma)$ at completion of iteration
numbers $1$,$3$ and $7$, together with the limiting equilibrium solution (solid
line). Also shown for comparison is the potential of mean force
$\mu^{ig}(\sigma)=\ln(\rho_t(\sigma))$. The upper part of the figure shows (on the same scale) a
difference plot measuring the deviation from the final solution at each
iteration.}
\label{fig:mu}
\end{figure}

Histograms of $\rho(\sigma)$ and $\mu(\sigma)$ were formed by
partitioning the range $0 \le \sigma \le \sigma_c$ into $M=120$ equal
intervals.  The refinement procedure was initialized by setting
$\mu_0(\sigma)=0\: \forall\: \sigma$. The initial modification factor
was set to $\beta_0=0.01$ (and halved at each successive iteration),
while the threshold value of $\zeta$ at which an iteration is deemed
complete (cf. Eq.~\ref{eq:zeta}) was set to $\zeta^*=0.4$. With these 
parameters the simulation was found to converge to the correct chemical
potential distribution in $7$ iterations ($\beta_7=1.56\times
10^{-4})$, requiring a total of approximately 100 hours CPU time on an
AMD $2000$+ processor. The form of $\mu(\sigma)$ at the $1$st, $3$rd
and $7$th iterations are compared with the limiting equilibrium form in
fig.~\ref{fig:mu}. One sees that the principal effect of the reduction
in $\beta$ at each successive iteration is a decrease in the scatter of
the solution; there seems to be no clear systematic dependence of the
solution on the value of $\beta$. This feature was confirmed by
independent runs at a number of other volume fractions and model
parameters.

\begin{figure}[h]
\includegraphics[width=8.0cm,clip=true]{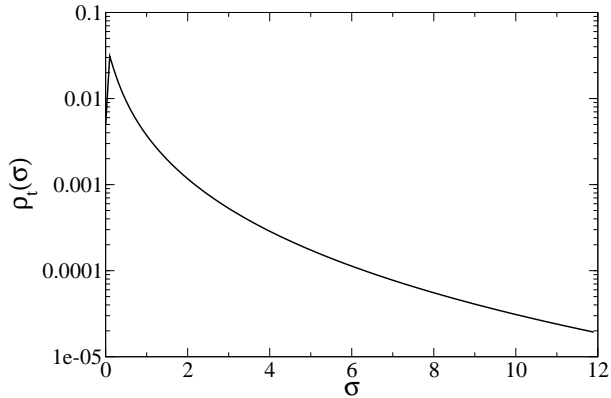}
\caption{The log-normal target distribution $\rho_t(\sigma)$ described
in the text and displayed on a logarithmic scale. The distribution has
standard deviation $W=2.5$ and scale parameter $\rho_t=0.0152$.}
\label{fig:lognorm}
\end{figure}

\begin{figure}[h]
\includegraphics[width=7.0cm,clip=true]{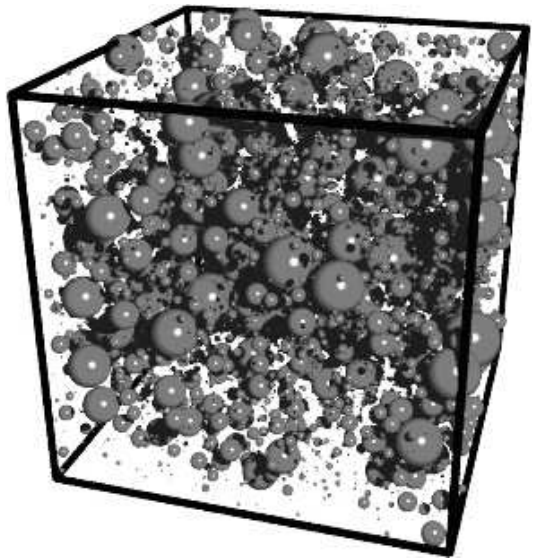}
\caption{Snapshot configuration of $N=11274$ polydisperse LJ particles
drawn from the log-normal size distribution shown in fig.~\protect\ref{fig:lognorm}.}
\label{fig:conf}
\end{figure}

Fig.~\ref{fig:conf} shows a snapshot of a typical configuration
obtained using the equilibrium solution for $\mu(\sigma)$. The
configuration contains some $1.1\times 10^4$ particles. Although the
form of $\rho_t(\sigma)$ (fig.~\ref{fig:lognorm}) dictates that the
largest particles occur with a probability of less than $0.5\%$ that of
the average particle size, their far greater volume implies that they
nevertheless occupy a sizeable fraction of the system volume. Herein
lies the physical origin of the  cross coupling between
$\overline\rho(\sigma)$ and $\mu(\sigma)$. Since the density (and
chemical potential) of the small particles depends on the available
volume, it must therefore also depend on the density (and chemical
potential) of the larger particles, and {\em vice-versa}.

Finally, for comparison, we have attempted to solve the same inverse
problem using the equilibrium scheme of eq.~\ref{eq:pmfa}. To
initialize the chemical potential distribution, we took $\mu_0(\sigma)=
\ln(\rho_t(\sigma))$. Some preliminary tuning of the damping factor
$\alpha$ and the run length of each iteration was found to be necessary
in order for the method to converge. Trials with $\alpha=1.0$,
$\alpha=0.5$ and $\alpha=0.25$ failed to converge in a reasonable
timescale. For $\alpha=0.1$ the method did converge (in an oscillatory
fashion), taking $93$ iterations to reach the solution (compared to $7$
iterations for the non-equilibrium method), and in so doing consuming
some $230$ hours of CPU time. A similar run, starting with
$\mu_0(\sigma)=0\: \forall\: \sigma$ (as used in the test of the
non-equilibrium approach) did not converge, thus demonstrating the
sensitivity of the method to the quality of the initial guess. For this
particular problem, the non-equilibrium scheme is therefore
computationally more efficient by at least a factor of two.

\section{Discussion and conclusions}

In this paper we have introduced a non-equilibrium iterative potential
refinement scheme for inverse problems. As a test, the method was used to
determine the chemical potential distribution for a LJ fluid having a
wide distribution of particle sizes. We find it to be efficient, simple to
implement and it requires no initial guess for the solution.

The efficacy of the new method arises from its application of
continuous small modifications to $\mu(\sigma)$ throughout the
refinement, rather than large-scale updates between iterations as
occurs in equilibrium schemes. The advantages of such an approach are
apparent when one considers the high degree of cross coupling that can
exist between $\mu(\sigma)$ and $\overline\rho(\sigma)$, and the
problem it implies: namely that a change to the chemical potential at
any one value of $\sigma$ can substantially affect the whole
composition distribution. Accordingly it is desirable that (a) each
individual modification to $\mu(\sigma)$ is kept small in order to
moderate the magnitude of cross coupling effects, and (b) such
deviations from the target distribution that do arise are immediately
corrected by further modifications to $\mu(\sigma)$. Our method
achieves this by continually `tweaking' $\mu(\sigma)$ by small amounts
to iron out differences between the {\em instantaneous} form of
$\rho(\sigma)$ and the target.

With regard to convergence properties, our findings indicate that even
for the earliest iterations, the method yields (at least for current
problem) an estimate for $\mu(\sigma)$ which, whilst statistically
poor, does not appear to deviate systematically from the true solution.
As iteration proceeds, it ``hugs'' the solution ever more tightly,
thereby allowing the degree of convergence to be judged solely on the
basis of the statistical scatter of the solution. This remarkable
feature would seem to promote reliable convergence to the true
solution. One might additionally speculate that the intrinsically
non-equilibrium character of the method generates ``noise'' that
promotes escape from any local minima in the solution space, thus
overcoming one of the apparent limitations of gradient-based
refinement schemes.  

The main factor that we have found to influence the reliability and
rate of convergence in our method is the choice of the initial value of
the modification factor $\beta_0$ \cite{NOTEZETA}. Should this value be
chosen to be too large ($\beta_0>0.05$) then the method may require an
excessive number of iterations to converge, or in extreme cases (as
also occurs in equilibrium schemes \cite{LYUBARTSEV95}), not converge
at all. If, on the other hand, it is chosen very small
($\beta_0<0.005$), then fewer iterations may be necessary overall, but
the earlier iterations will take longer to complete because they
require many more modifications steps. Within this window, the
efficiency does not appear to vary by more than about a factor of $2$,
but there is certainly scope for trial and error optimization. We found
that this is achievable quickly (during the initial stages of the first
iteration) simply by monitoring the time evolution of $\zeta$ (c.f.
eq.~\ref{eq:zeta}). In cases where $\beta_0$ is chosen too large for
convergence, this quickly becomes evident during the first iteration
because $\zeta$ does not decrease steadily towards $\zeta^\star$. This
permits early diagnosis of convergence problems, in contrast to
equilibrium schemes, where a failure to converge is typically only
apparent after a number of iterations.

Finally we note that although we have tested our method within the
specific context of a polydisperse LJ fluid, its potential
applicability is more wide ranging. As described in
sec.~\ref{sec:intro}, a diverse range of inverse problems can be mapped
onto a grand canonical polydisperse composition space \cite{RUT01}. It
is simply a matter of identifying the analogue quantities to
$\rho(\sigma)$ and $\mu(\sigma)$; thereafter the method can be applied
exactly as we have described. Thus, for instance, in a molecular
liquid, a (partial) structure factor $g(r)$ might play the role of
$\rho(\sigma)$ while the interatomic pair potential $\phi(r)$ plays the
role of $\mu(\sigma)$. If applied to real experimental data in this
way, it would be of interest to compare the robustness of our method
with that of existing refinement schemes, particularly with regard to
the sensitivity to experimental uncertainty in the target structural
data and potential truncation effects \cite{SOPER96}.

\subsection*{Acknowledgment}

The author is grateful P. Sollich for discussions and to A.D. Bruce for
a thought provoking murmur. This work was supported by the EPSRC, grant
number GR/S59208/01 and the Royal Society.



\appendix

\section{Grand canonical formulation of polydispersity}
\label{append}
Here we provide a brief formal overview of the statistical mechanics 
of polydispersity. We consider a classical fluid of polydisperse particles confined to a
volume $V=L^d$. The system is assumed to be thermodynamically open, so
that the particle-number distribution $N(\sigma)$ is a statistical
quantity. The associated grand canonical partition function takes the
form:

\begin{equation} 
\label{eq:bigzdef} 
{\cal Z}_V  = \sum _{N=0}^{\infty }\frac{1}{N!}\prod_{i=1}^{N} \left\{\int_V d\vec{r}_i\int_0^\infty d\sigma_i\right\} \exp{\left(-\beta{\cal H}_N\left(\{\vec{r},\sigma\}\right)\right)}
\end{equation} 
with
\begin{equation}
{\cal H}_N\left(\{\vec{r},\sigma\}\right)=\Phi \left(\{ \vec{r},\sigma \}\right)-\sum_{i=1}^N\mu(\sigma_i)\;.
\end{equation}
Here $N$ is the overall particle number, while $\beta=(k_BT)^{-1}$  and
$\mu(\sigma)$ are respectively the prescribed inverse temperature and
chemical potential distribution. $\{\vec{r},\sigma\}$ denotes the {\em
configuration}, i.e. the complete set $(\vec{r_1},\sigma_1),(\vec{r_2},
\sigma_2)\cdots (\vec{r_N},\sigma_N)$ of particle position vectors and
polydisperse attributes. The corresponding configurational energy $\Phi
\left(\{ \vec{r},\sigma \}\right)$ is assumed to reside in a sum of
pairwise interactions

\begin{equation}
\Phi \left(\{ \vec{r},\sigma \}\right)= \sum_{i<j =1}^N\phi(\vec{r}_i,\vec{r}_j,\sigma_i,\sigma_j)\;,
\end{equation}
where $\phi$ is the pair potential.

The fluctuating particle number distribution is defined by

\begin{equation}
N(\sigma) \equiv \sum_{i=1}^N \delta(\sigma-\sigma_i)\;,
\end{equation}
with $\sigma$ the continuous polydispersity attribute and $N=\int
N(\sigma)d\sigma$. The associated composition distribution is

\begin{equation}
\rho(\sigma)\equiv N(\sigma)/V\;.
\end{equation}

The statistical behavior of $\rho(\sigma)$ is completely described
by its probability distribution

\begin{eqnarray}
p_V[\rho(\sigma)]\hspace*{-1mm} &=& \hspace*{-1mm}\frac{1}{{\cal Z}_V}\sum_{N=0}^\infty\frac{1}{N!}\prod _{i=1}^{N} \left\{\int_V d\vec{r}_i\int_0^\infty d\sigma_i\right\}  \\ \nonumber
 \times &\;& \hspace*{-5mm} \exp{\left(-\beta{\cal H}_N\left(\{\vec{r},\sigma\}\right)\right )} \prod_\sigma\delta \left( \rho(\sigma) - N(\sigma)/V \right) \;,
\label{eq:jointdist}
\end{eqnarray}
from which the ensemble averaged form of $\rho(\sigma)$ follows as

\begin{equation}
\overline{\rho}(\sigma)=\int\rho(\sigma)p_V[\rho(\sigma)]d\rho(\sigma)\;.
\end{equation}


\begin{thebibliography}{99}
\bibitem{NOTE0} An equivalent viewpoint (adopted in \protect\cite{RUT01}) 
considers $N$ pseudo particles speciated according to their position
relative to the rest of the system. The species label is then formally
an $N$-component vector $\vec\sigma$ corresponding to the $N$
interparticle distances.  While this formulation obviates the need for
a configurational constraint, the fact that $\mu(\vec\sigma)$ is
defined on an $N$-dimensional space, renders it computationally
cumbersome.
\bibitem{NOTECP} Here and throughout this paper, factors of the inverse
temperature $\beta$ have been absorbed into the effective chemical
potential distribution.
\bibitem{NOTE1} Although we have no definitive explanation for this
observation, it seems likely that occasional iterations of the form
Eq.~\ref{eq:pmfa} introduce noise into the minimization procedure,
thereby enabling escape from local minima.
\bibitem{NOTEZETA} In principle, convergence might also fail  if one
were to choose an excessively large value of $\zeta^\star$. However,
the value $\zeta^\star=0.4$ we suggest here works well in all cases we
have examined.
\end{thebibliography}
\end{document}